\documentclass[twocolumn]{aastex701}

\usepackage{amsmath}

\begin{document}

\title{Extending the micro-Hertz detection horizons via orbital resonance effect for geocentric gravitational wave antennas}

\author[orcid=0009-0005-8281-0238]{Qiong Deng}
\altaffiliation{Corresponding authors}
\affiliation{Lanzhou Center for Theoretical Physics, Key Laboratory of Theoretical Physics of Gansu Province, and Key Laboratory of Quantum Theory and Applications of MoE, Lanzhou University, Lanzhou, 730000, China}
\affiliation{Center for Gravitational Wave Experiment, National Microgravity Laboratory, Institute of Mechanics, Chinese Academy of Sciences, Beijing, 100190, China}
\email[show]{Dengq19@lzu.edu.cn}  

\author[orcid=0000-0003-2155-3280]{Minghui Du} 
\affiliation{Center for Gravitational Wave Experiment, National Microgravity Laboratory, Institute of Mechanics, Chinese Academy of Sciences, Beijing, 100190, China}
\email{duminghui@imech.ac.cn}

\author[orcid=0000-0002-3543-7777]{Peng Xu}
\altaffiliation{Corresponding authors}
\affiliation{Center for Gravitational Wave Experiment, National Microgravity Laboratory, Institute of Mechanics, Chinese Academy of Sciences, Beijing, 100190, China}

\affiliation{Lanzhou Center for Theoretical Physics, Key Laboratory of Theoretical Physics of Gansu Province, and Key Laboratory of Quantum Theory and Applications of MoE, Lanzhou University, Lanzhou, 730000, China}
                        
\affiliation{Taiji Laboratory for Gravitational Wave Universe (Beijing/Hangzhou), University of Chinese Academy of Sciences, Beijing, 100049, China}
            
\affiliation{Hangzhou Institute for Advanced Study, University of Chinese Academy of Sciences, Hangzhou, 310124, China}
\email[show]{xupeng@imech.ac.cn}

\author{Liang Huang}
\affiliation{Lanzhou Center for Theoretical Physics, Key Laboratory of Theoretical Physics of Gansu Province, and Key Laboratory of Quantum Theory and Applications of MoE, Lanzhou University, Lanzhou, 730000, China}%
\email{huangl@lzu.edu.cn}

\author[0000-0002-9533-8025]{Ziren Luo}
\affiliation{Center for Gravitational Wave Experiment, National Microgravity Laboratory, Institute of Mechanics, Chinese Academy of Sciences, Beijing, 100190, China}
            
\affiliation{Taiji Laboratory for Gravitational Wave Universe (Beijing/Hangzhou), University of Chinese Academy of Sciences, Beijing, 100049, China}
            
\affiliation{Hangzhou Institute for Advanced Study, University of Chinese Academy of Sciences, Hangzhou, 310124, China}
\email{luoziren@imech.ac.cn}




\begin{abstract}

The $\mu$Hz gravitational wave band holds crucial insights into coalescing supermassive black hole binaries and stochastic backgrounds but remains inaccessible due to technical challenges. We demonstrate that geocentric space-based GW detectors (e.g., TianQin, gLISA, GADFLI) can bridge this gap by considering orbital resonance effects, circumventing the need for prohibitively long baselines. When GW frequencies match with integer multiples of a satellite’s orbital frequency, sustained tidal forces induce cumulative orbital deviations through resonant effects, which, combined with orbital modulation, improve detector sensitivity by 1–2 orders of magnitude in the $\mu$Hz band. Consequently, geocentric missions can detect SMBHBs across significantly expanded mass-redshift parameter space. Crucially, such observations could synergize with pulsar timing array data of the same binaries at earlier inspiral stages, enabling unprecedented joint tests of strong-field gravity and binary evolution. Our findings establish geocentric antennas as a cost-effective, near-term precursor for unlocking the $\mu$Hz GW astronomy.

\end{abstract}



\section{Introduction} 

In 2015, LIGO’s first direct detection of gravitational waves (GWs) marked a pivotal moment in astronomy and cosmology, opening a new era in unprecedented probing of strong-field gravity, compact objects and exploring the Universe \citep{abbott2016observation}. 
Current and planned GW observatories are optimized for distinct frequency bands for specific scientific objectives.
Pulsar Timing Arrays (PTAs) \citep{falxa2023searching,xu2023searching,epta2023second,epta2024second} access the nanohertz regime, space-based antennas including LISA \citep{colpi2024lisa}, Taiji \citep{luo2021taiji}, TianQin \citep{luo2016TianQin,luo2025progress, huang2020science} and Decigo \citep{kawamura2006japanese} will cover the 0.1 mHz to 1 Hz band, and ground-based detectors LIGO–Virgo–KAGRA (LVK) collaboration \citep{abbott2009ligo,acernese2014advanced,akutsu2021overview,collaboration2019advanced} and planned 3rd-generation detectors \citep{punturo2014third,reitze2019cosmic} mainly operate above $\sim$ 10 Hz. 
These, together with the CMB observations (1 aHz - 10 fHz) \citep{abazajian2022cmb,li2019probing,ghosh2022performance,bucher2017physics},  span an extremely wide spectrum of the GW landscape.

Crucially, the $\mu$Hz window ($10^{-7}$–$10^{-4}$ Hz), lying between the sensitivity ranges of PTA and space-based antennas, remains still observationally challenging. 
This gap obscures key astrophysical processes, including the final coalescence of supermassive black hole binaries (SMBHBs) with total masses  $10^7$-$10^{10}\,M_\odot$ (cosmic engines driving galaxy evolution), and potential relic stochastic backgrounds from the early Universe. 
A growing sample of candidate SMBHBs has been identified via multiwavelength surveys (e.g., OJ 287 \citep{valtonen2008massive}, PG 1302–102 \citep{graham2015possible}, Mrk 231 \citep{yan2015probable}, B2 0402+379 \citep{bingol2006activity}, MCG +11–11–032 \citep{ardaneh2018direct}, SDSSJ1430+2303 \citep{jiang2022tick}), yet their GW signatures remain undetectable with current instruments.  
Detecting these systems would help resolve SMBHB formation channels, trace black hole growth, and test general relativity in extreme strong-field regimes. 

Bridging the $\mu$Hz window has motivated the development of dedicated detection strategies. Proposed solutions, such as space interferometers with baseline $\sim 1AU$ including $\mu$Ares \citep{sesana2021unveiling}, LISAmax \citep{martens2023lisamax}, ASTROD-GW \citep{ni2013astrod}, and Super-ASTROD \citep{ni2009super}, face prohibitive technological and financial challenges.
Here, we demonstrate that the existing \textit{geocentric} GW antenna concepts, including LISA-like missions (LISA-like measurement concept and payload architecture) gLISA \citep{2015gLISA, tinto2011geostationary}, TianQin \citep{luo2016TianQin,luo2025progress, huang2020science}, OMEGA \citep{hellings2011low}, LAGRANGE \citep{conklin2011lagrange}, ASTROD-EM \citep{ni2016gravitational}, and GADFLI \citep{mcwilliams2011geostationary}, which are designed for $\mu$Hz to deciHertz science but orbiting Earth with $\mu$Hz orbital frequencies, can unlock this band via a long-overlooked physical mechanism, that the \textit{orbital resonant effect}. 
Early studies had demonstrated that when incident GWs frequencies match the integer multiples of the orbital frequency of the satellite, sustained tidal forces from the GWs could drive cumulative orbital deviations \citep{mashhoon1978tidal, rudenko1975test, turner1979influence},  which could therefore amplify the detector responses. 
Recent advances in space-based precision metrology have renewed interest in this approach, especially for feasibility studies for Satellite Laser Ranging (SLR) missions \citep{blas2022bridging, blas2022detecting, 2023Probing,foster2025discovering}. 
In particular, the seminal works by Blas and Jenkins \citep{blas2022bridging, blas2022detecting, foster2025discovering} established a comprehensive theoretical and numerical framework, highlighting the potential of the practical use of such measurement scheme for probing the $\mu$Hz GW band. 

Motivated by these results \citep{blas2022bridging, blas2022detecting, foster2025discovering}, in this work, we focus on the enhancements in geocentric interferometric antennas from orbit resonance and orbital modulations. Our analysis shows that such effects could enhance the $\mu$Hz sensitivity of geocentric antennas by 1–2 orders of magnitude, extend the detectable horizon for coalescing SMBHBs and increase the observable population compared to conventional sensitivity assessments, and therefore enable missions like TianQin, gLISA, and GADFLI to track individual SMBHBs through inspiral-merger-ringdown evolution. Crucially, resonant $\mu$Hz detections provide possible complementary data on late-stage SMBHB dynamics, which may synergize with PTA observations of the same systems years to decades earlier in their nHz inspiral phase and enable unprecedented joint tests of binary evolution, environmental coupling, and strong-field gravity across cosmic time. 
This resonance-driven approach establishes geocentric antennas as a cost-effective, lower-risk precursors to bridge the $\mu$Hz gap, and together with SLR and Lunar laser ranging missions, a coordinated network for $\mu$Hz GW observation could be formed.

\section{Detection scheme} \label{scheme}
\begin{figure}[ht]
	\centering 
	\includegraphics[width=0.4\textwidth, angle=0]{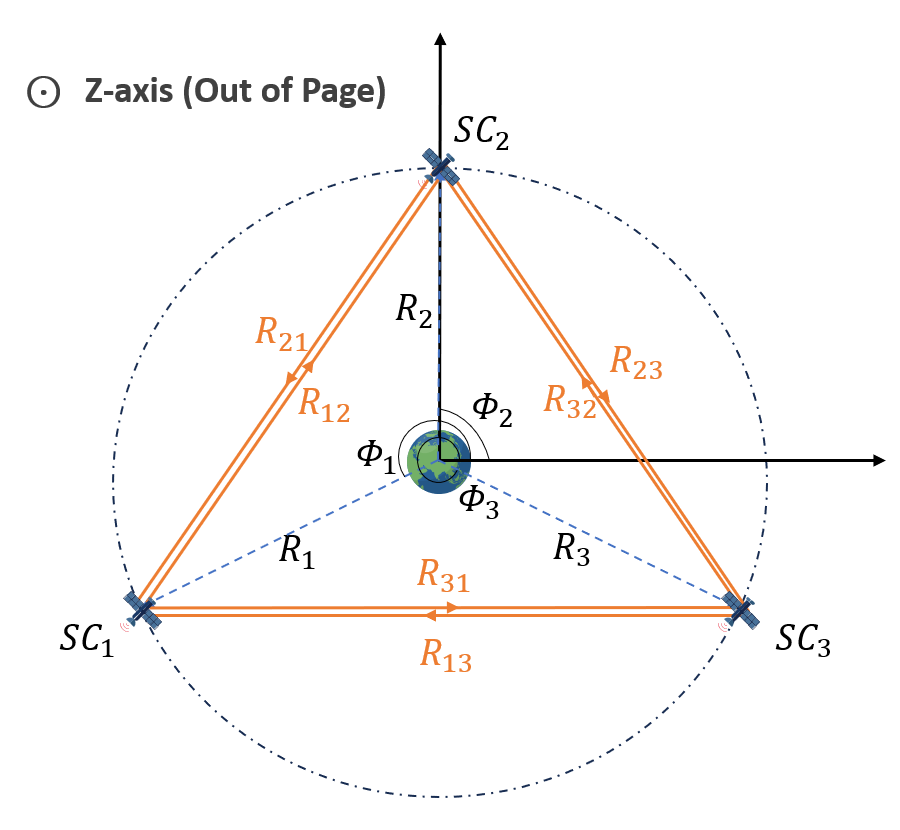}
	\caption{A simplified constellation configuration to model geocentric GW space antennas. Three SCs co-orbit Earth in a common circular orbit, forming an equilateral triangle, maintaining laser interferometry links between each pair and effectively forming multiple partially independent Michelson-type interferometers.} 
	\label{fig：config}
\end{figure}

To simplify the analysis, we model geocentric space GW antennas using the constellation configuration in Figure~\ref{fig：config}: three spacecrafts (SCs) co-orbit Earth in a common circular orbit, forming an equilateral triangle, maintaining laser interferometry links between each pair and effectively forming multiple partially independent Michelson-type or Sagnac-type interferometers. Under the long-wavelength approximation (GW wavelength $\gg$ SC-Earth distance), the effect of GWs can be modeled as perturbing tidal forces acting on the system \citep{maggiore2007gravitational}. This simplifies the problem to a classical perturbed two-body problem, an approach used in previous studies \citep{mashhoon1978tidal,rudenko1975test,turner1979influence,blas2022detecting}. For the subsequent analysis, we adopt the perturbation formalism developed by Mashhoon \citep{mashhoon1978tidal} rather than the more recent formulations \citep{blas2022bridging, blas2022detecting, foster2025discovering}, as it allows a more direct derivation of the analytical form of the resonant orbital response, which is central to our study. This approach provides a unified theoretical framework that enables a consistent comparison of the orbital resonance effect across different proposed geocentric missions, despite their different orbital configurations and instrumental designs. By working within a simplified analytical model, that Keplerian two-body systems with initially circular orbits and a static triangular constellation, we establish a well-defined baseline. This idealized setup isolates the intrinsic sensitivity enhancement due to the resonance mechanism alone, decoupling it from mission-specific complexities.

We employ a cylindrical coordinate system $(r, \phi, z)$ with its origin located at the system's center of mass, the $r$-$\phi$ plane coincides with the orbital plane, and the z-axis is normal to it. The coordinates of each SC under GW perturbations are denoted by uppercase letters $(R_n, \Phi_n, Z_n)$, while lowercase letters $(r_n, \phi_n, z_n)$ represent their unperturbed positions, where \( n = 1, 2, 3 \) labels the \( n \)-th SC in the detector constellation. 
The perturbed orbits $(R_n(t),\Phi_n(t),Z_n(t))$ could be expanded as
\begin{align}
		R_n(t)&=r_n(t)+r_n(t)\epsilon F_n(t)+O(\epsilon^2) ,\label{eq:3}\\
        \Phi_n(t)&=\phi_n(t)+\epsilon G_n(t)+O(\epsilon^2) \label{eq:4},\\
        Z_n(t)&=r_n(t)\epsilon H_n(t)+O(\epsilon^2),
\end{align}
where $\{F_n(t), G_n(t), H_n(t)\}$ are the first-order GW perturbations of the orbit, the small parameter $\epsilon\sim |h|$ is used here to represent the strength of the incident GW. With the simplified model, the unperturbed circular orbits satisfy $r_n=r_0$ and $\phi_n(t)=\omega_0 t+\phi_{0,n}$, where $r_0$ is the orbital radius, $\omega_0$  the angular frequency, and $\phi_{0,n}$ the initial orbital phase of the n-th SC. Following the definition in \citep{cornish2001detecting}, the readout of the Michelson-type interferometer from $SC_i$ can be written down as
\begin{align}
    M_{i}(t)=&\frac{r_0 \epsilon}{2L}[\sqrt{3}(F_j(t)-F_k(t))\nonumber\\
    &+(G_j(t)+G_k(t)-2G_i(t))]+O(\epsilon^2)\label{ep:7}.
\end{align}
For any given $SC_i$, the indices of its two adjacent SCs are defined as $j = (i + 1) \bmod 3$, $k = (i + 2) \bmod 3$. The initial orbital phase differences are assumed to satisfy $\phi_{0,i}-\phi_{0,j}=\phi_{0,k}-\phi_{0,i}={2\pi}/{3}$, since the idealized equilateral triangular constellation. $L$ is the nominal distance between any two SCs. In particular, Eq. \eqref{ep:7} shows that the impact of $H_n$ on the observable $M$ is of the order $O(\epsilon^2)$. Therefore, perturbation in the z-direction is considered to be too small and can be ignored. Substituting Eqs.~\eqref{eq:3}-\eqref{eq:4} into the perturbation equations of Keplerian orbital motion and retaining terms up to first order in $\epsilon$, the linearized dynamics for radial ($F_n$) and azimuthal ($G_n$) perturbations satisfy
\begin{align}
		\ddot{F}_n(t)-2\omega_0\dot{G}_n(t)-3\omega_0^2 F_n(t)=\frac{f_{R,n}(t)}{\epsilon r_0}+O(\epsilon^2), \label{eq:5}\\
		\ddot{G}_n(t)+2\omega_0\dot{F}_n(t)=\frac{f_{\Phi,n}(t)}{\epsilon r_0}+O(\epsilon^2)\label{eq:6}.
\end{align}
Here, $f_{R,n}(t)$ and $f_{\Phi,n}(t)$ are the radial and azimuthal components of the GW tidal force per unit mass. Equations~\eqref{eq:5} and~\eqref{eq:6} are derived under the assumption of circular orbits. In fact, the detector configurations discussed in this paper are designed to operate on nearly circular orbits with small eccentricities. This design choice is not incidental: because these missions rely on stable inter-spacecraft laser links, the breathing angles (i.e., the variations in the angles between the constellation's arms) of the constellation must remain small. Maintaining long-term link stability and avoiding excessive variations in the arm lengths requires SCs to follow nearly circular orbits. When the eccentricity is sufficiently small, the orbital motion can be expanded in powers of the eccentricity, where the zeroth-order term corresponds to the circular-orbit solution. As shown in Appendix \ref{AppendixA}, the circular-orbit solution provides the dominant contribution in the small-eccentricity limit.
In the frequency domain, the solutions of Eq. \eqref{eq:5} and Eq. \eqref{eq:6} can be derived
\begin{align}   
\tilde{F}_n(\omega)&=\frac{\omega^2\tilde{f}_{R,n}(\omega)-2i\omega_0\omega\tilde{f}_{\Phi,n}(\omega)}{\epsilon r_0\omega^2(\omega_0^2-\omega^2)},\label{eq:10}\\
    \tilde{G}_n(\omega)&=\frac{2i\omega\omega_0\tilde{f}_{R,n}(\omega)+(\omega^2+3\omega_0^2)\tilde{f}_{\Phi,n}(\omega)}{\epsilon r_0\omega^2(\omega_0^2-\omega^2)}.\label{eq:11}
\end{align}
The symbol $\tilde{\ }$ denotes the Fourier transform of the corresponding physical quantity. 

The tidal force induced by GW is
\begin{align}
	f_l(t)=\frac{1}{2}\ddot{h}_{lm}(t)x^m=\frac{1}{2}e^A_{lm}\ddot{h}_A x^m,
\end{align}
where $h_{lm}$ is the transverse-traceless part of the metric
perturbation, $A=+,\times$ denotes the two GW polarizations, $e^A_{lm}$ be the standard polarization tensor, and the vector $\vec{x}\equiv(r\cos{\phi}, r\sin{\phi}, z)^T$.
So the radial and azimuthal components of the GW tidal force per unit mass in the frequency domain read
\begin{align}
	\tilde{f}_{R_n}&(\omega)=-\frac{r_0}{2}[\frac{e^A_{11}+e^A_{22}}{2} \tilde{h}_A(\omega)+\frac{(e^A_{11}-e^A_{22})/2-ie^A_{12}}{2}\notag\\ \times&(\omega-2\omega_0)^2e^{i2\phi_{0,n}}\tilde{h}_A(\omega-2\omega_0)+\frac{(e^A_{11}-e^A_{22})/2+ie^A_{12}}{2}\notag\\ \times&(\omega+2\omega_0)^2 e^{-i2\phi_{0,n}}\tilde{h}_A(\omega+2\omega_0) ] \label{eq:8},\\
	\tilde{f}_{\Phi_n}&(\omega)=-\frac{ir_0}{2}[\frac{(e^A_{11}-e^A_{22})/2-ie^A_{12}}{2} (\omega-2\omega_0)^2e^{i2\phi_{0,n}}\notag\\\times&\tilde{h}_A(\omega-2\omega_0)-\frac{(e^A_{11}-e^A_{22})/2+ie^A_{12}}{2}(\omega+2\omega_0)^2 e^{-i2\phi_{0,n}}\notag\\\times&\tilde{h}_A(\omega+2\omega_0)]\label{eq:9}.
\end{align}
The denominators in Eqs.~\eqref{eq:10}-\eqref{eq:11} reveal two distinct resonant conditions: one at \(\omega = 0\) and the other at \(\omega = \omega_0\). 
Analysis of the driving forces in Eqs.~\eqref{eq:8}-\eqref{eq:9} shows:
\begin{itemize}
    \item The \(\omega = 0\) resonance is excited by GW components at frequencies \(\tilde{h}_A(0)\), \(\tilde{h}_A(2\omega_0)\), and \(\tilde{h}_A(-2\omega_0)\),
    \item The \(\omega = \omega_0\) resonance is excited by GW components at \(\tilde{h}_A(\omega_0)\), \(\tilde{h}_A(-\omega_0)\), and \(\tilde{h}_A(3\omega_0)\).
\end{itemize}
This \(2\omega_0\)  frequency shift between incident GW signal and the GW driving force originates from the frequency modulation due to the SC’s periodic orbital motion.  

\begin{figure}[ht]
	\centering 
	\includegraphics[width=0.45\textwidth]{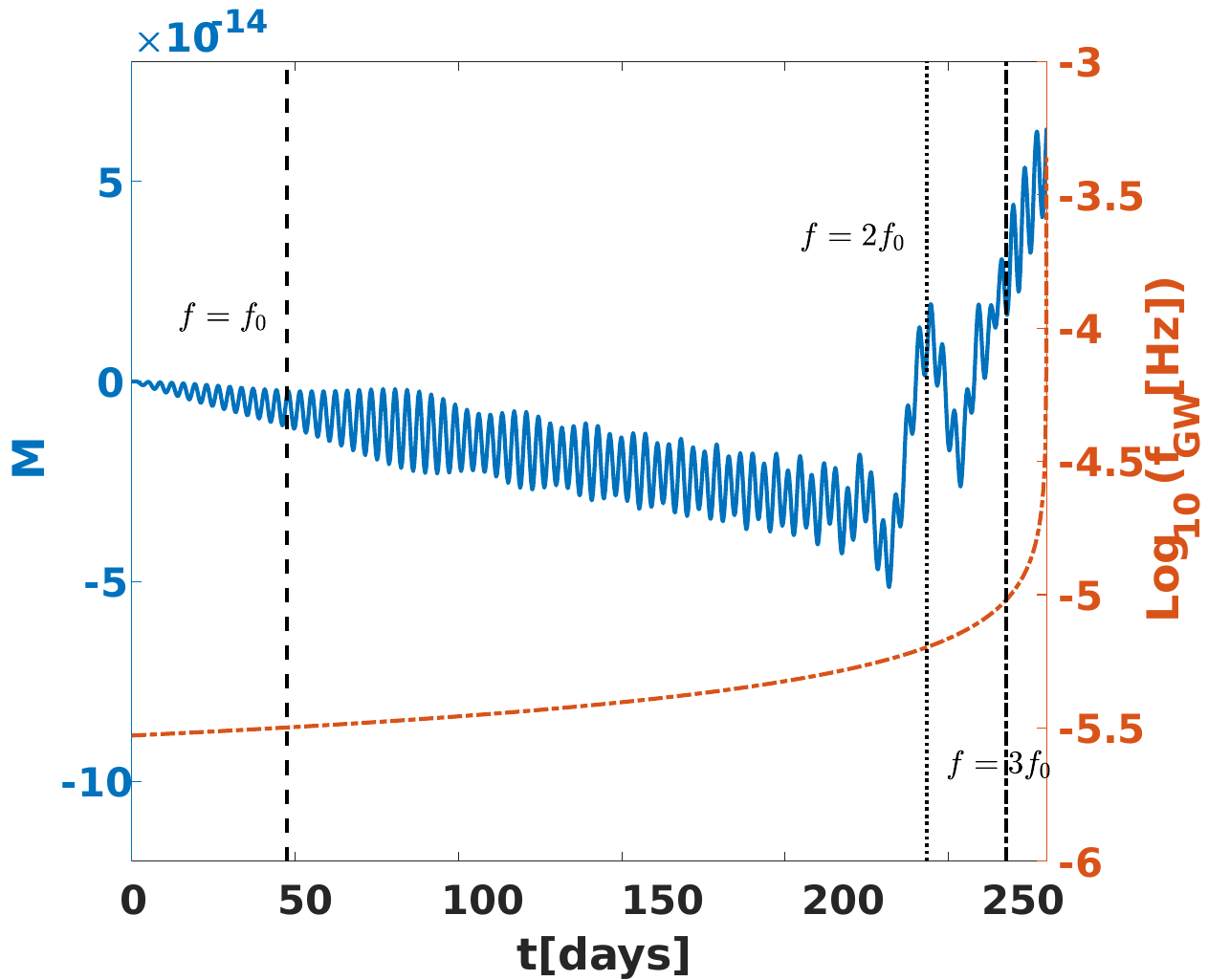}
	\caption{The Michelson-type response $M$ (blue curve) to a coalescing SMBHB signal as its frequency (red curve) evolves. When the GW frequency crosses $f_0$ and $3f_0$, $M$ grows linearly with time, while it crosses $2f_0$ M exhibits a quadratic ($\sim t^2$) behavior. }
	\label{fig：1}
\end{figure}

Consequently, GWs at three distinct frequencies, $\omega_0$, $2\omega_0$, and $3\omega_0$, can drive the circular orbit binary system to resonance. This behavior is fully consistent with Mashhoon's theoretical analysis \citep{mashhoon1978tidal}, with the following characteristic time-domain responses:
\begin{itemize}
    \item For $\omega_{\rm GW} = 2\omega_0$: A constant azimuthal force induces a quadratic secular change in orbital phase, 
    \item For $\omega_{\rm GW} = \omega_0$ or $3\omega_0$: Both radial and azimuthal motions exhibit periodic perturbations with amplitudes growing linearly in time. 
\end{itemize}
Detailed analytical solutions for these time-domain responses are provided in \citep{mashhoon1978tidal}. Figure~\ref{fig：1} shows, as a representative example, the response of a geocentric GW antenna (with $r_0 = 10^5$ km) to a coalescing SMBHB signal (total mass $8 \times 10^7 M_{\odot}$). The GW propagates in the direction perpendicular to the antenna's orbital plane. 
When the frequency crosses $f_0$ and $3f_0$, the response amplitude increases linearly over time; when it crosses $2f_0$, it exhibits the quadratic ($M_i \sim t^2$) behavior. These sequential resonances produce both the amplifications and characteristic fingerprints of the observable $M_i$, enabling precision estimations of source parameters against the instrumental noises and non-GW disturbances in the $\mu$Hz band.

\section{Sensitivity analysis} 
To compute the sensitivity of the detector, one can define the sky and polarization averaged signal response function of the detector $\mathcal{R}(\omega,\omega_0)$ as 
\begin{align}
\mathcal{R}&(\omega,\omega_0)\tilde{h}^*(\omega)\tilde{h}(\omega)
    \notag\\&\equiv \frac{1}{4\pi^2}\int_0^\pi \rm{d}\psi\int_0^{2\pi} \rm{d}\varphi\int_0^{\pi} \tilde{M}_i^*(\omega)\tilde{M}_i(\omega) \sin\theta\rm{d}\theta\notag\\&=\frac{1}{10\sqrt{3}}\zeta_{-}^*(\omega)\zeta_{-}(\omega)\tilde{h}^*(\omega-2\omega_0)\tilde{h}(\omega-2\omega_0)\notag\\&+\frac{1}{10\sqrt{3}}\zeta_{+}^*(\omega)\zeta_{+}(\omega)\tilde{h}^*(\omega+2\omega_0)\tilde{h}(\omega+2\omega_0),\label{eq:14}
\end{align}
where ($\theta,\varphi$) defines the sky location of the source, $\psi$ is the polarization angle, and
\begin{align}
    \zeta_{-}&=(\omega-2\omega_0)^2[ \frac{\sqrt{3}(\omega-2\omega_0)(e^{i2\phi_{0,j}}-e^{i2\phi_{0,k}})}{4\omega(\omega^2-\omega_0^2)}\notag\\ +&\frac{i(\omega^2+3\omega_0^2-2\omega\omega_0)(e^{i2\phi_{0,j}}+e^{i2\phi_{0,k}}-2e^{i2\phi_{0,i}})}{4 \omega^2(\omega^2-\omega_0^2)}], \\
    \zeta_{+}&=(\omega+2\omega_0)^2[ \frac{\sqrt{3}(\omega+2\omega_0)(e^{i2\phi_{0,j}}-e^{i2\phi_{0,k}})}{4\omega(\omega^2-\omega_0^2)}\notag\\ -&\frac{i(\omega^2+3\omega_0^2+2\omega\omega_0)(e^{i2\phi_{0,j}}+e^{i2\phi_{0,k}}-2e^{i2\phi_{0,i}})}{4 \omega^2(\omega^2-\omega_0^2)}],
\end{align}
\begin{align}
    \tilde{h}^*(\omega)\tilde{h}(\omega)\equiv\tilde{h}^*_+(\omega)\tilde{h}_+(\omega)+\tilde{h}^*_\times(\omega)\tilde{h}_\times(\omega).
\end{align}

As evident from Eq.~\eqref{eq:14}, the precise form of the response function $\mathcal{R}(\omega,\omega_0)$ is directly determined by the specific spectral properties of the GW signal $\tilde{h}(\omega)$. For the class of coalescing SMBHBs considered in this work, we adopt the phenomenological inspiral-merger-ringdown (IMR) model PhenomA \citep{ajith2007phenomenological} as our fiducial waveform. Crucially, within this model, the dominant frequency dependence during the inspiral and merger regimes can be characterized by distinct, characteristic power-laws. This allows us to treat these evolutionary stages separately and derive approximate, regime-dependent response functions.

For the inspiral phase, the frequency evolution follows the post-Newtonian quadrupole formula, leading to the relation
\begin{align}
\tilde{h}^*(\omega-2\omega_0)\tilde{h}(\omega-2\omega_0)=\left(\frac{\omega-2\omega_0}{\omega}\right)^{-7/3}\tilde{h}^*(\omega)\tilde{h}(\omega),\notag\\
\tilde{h}^*(\omega+2\omega_0)\tilde{h}(\omega+2\omega_0)=\left(\frac{\omega+2\omega_0}{\omega}\right)^{-7/3}\tilde{h}^*(\omega)\tilde{h}(\omega).
\end{align}
Substituting this into Eq.~\eqref{eq:14} yields the response function for the inspiral phase:
\begin{align}
\mathcal{R}_{\mathrm{insp}}(\omega,\omega_0)=\frac{1}{10\sqrt{3}}\zeta_{-}^*(\omega)\zeta_{-}(\omega)\left(\frac{\omega-2\omega_0}{\omega}\right)^{-7/3}\notag\\+\frac{1}{10\sqrt{3}}\zeta_{+}^*(\omega)\zeta_{+}(\omega)\left(\frac{\omega+2\omega_0}{\omega}\right)^{-7/3}.\label{eq:16}
\end{align}

For the merger phase, phenomenological models indicate a steeper frequency scaling. Adopting the characteristic $\omega^{-4/3}$ dependence from the PhenomA model \citep{ajith2007phenomenological}, the corresponding relations are
\begin{align}
\tilde{h}^*(\omega-2\omega_0)\tilde{h}(\omega-2\omega_0)=\left(\frac{\omega-2\omega_0}{\omega}\right)^{-4/3}\tilde{h}^*(\omega)\tilde{h}(\omega),\notag\\
\tilde{h}^*(\omega+2\omega_0)\tilde{h}(\omega+2\omega_0)=\left(\frac{\omega+2\omega_0}{\omega}\right)^{-4/3}\tilde{h}^*(\omega)\tilde{h}(\omega).
\end{align}
The response function for the merger phase is therefore
\begin{align}
\mathcal{R}_{\mathrm{merg}}(\omega,\omega_0)=\frac{1}{10\sqrt{3}}\zeta_{-}^*(\omega)\zeta_{-}(\omega)\left(\frac{\omega-2\omega_0}{\omega}\right)^{-4/3}\notag\\+\frac{1}{10\sqrt{3}}\zeta_{+}^*(\omega)\zeta_{+}(\omega)\left(\frac{\omega+2\omega_0}{\omega}\right)^{-4/3}.\label{eq:17}
\end{align}

The ringdown stage is omitted from this study for two reasons. First, its response depends strongly on the specific parameters of the source, which prevents the derivation of a generic analytical form. Second, the ringdown signal's contribution to the SNR enhancement via orbital resonance is negligible.

These response functions, $\mathcal{R}_{\mathrm{insp}}$ and $\mathcal{R}_{\mathrm{merg}}$, provide the foundation for constructing the generic sensitivity curves used to compare different detector designs in the following sections.

The definition of the sensitivity curve follows that in \citep{robson2019construction}, as 
\begin{align}
    S_n(f)\equiv \frac{P_n(f)}{\mathcal{R}(f)}.
\end{align}
$S_n(f)$ is konwn as the PSD sensitivity. The Characteristic strain is defined as \cite{moore2014gravitational}
\begin{align}
    h_c(f)\equiv \sqrt{fS_n(f)}.\label{eq:18}
\end{align}
The form of $\mathcal{R}(f)$ is given by Eq. \eqref{eq:16}, and the total noise in a Michelson-type data channel $P_n(f)$ of a LISA-like mission is based on the following expression, see \citep{robson2019construction,cornish2001detecting}
\begin{align}
P_n(f)=\frac{P_{\text{OMS}}}{L^2}+2(1+\cos^2{f/f_*})\frac{P_{\text{ACC}}}{(2\pi f)^4L^2},\label{eq:17}
\end{align}
where $f_*=c/(2\pi L)$, $P_{\text{OMS}}$ is the one-way single-link optical metrology noise with laser frequency noises and clock noises removed, and $P_{\text{ACC}}$ the residual acceleration noise of the gravitational reference sensor. One notices that precision GW extractions for LISA-like missions require sophisticated data analysis techniques, such as data preprocessing techniques, source parameter inversions based on matched filtering \citep{christensen2022parameter, meyer2022computational} or deep learning techniques \citep{cuoco2020enhancing}, and so on. Since such data analysis procedures are unrelated to the resonance dynamics, and to ensure a unified and concise treatment of geocentric LISA-like missions, in this work we adopt the simplified Michelson-type observable given in Eq. (\ref{ep:7}), along with its total noise as defined in the preceding equation. For the time delay interferometry (TDI) channels for laser frequency noise suppressions \citep{tinto2021time}, the first-generation Michelson-type TDI channel $X$ for all detectors under the equal-armlength approximation is adopted here, with the response function and noise power spectral density given by,
\begin{align}
\mathcal{R}^X(f) = (1-\tilde{D}^2)\mathcal{R}(f),\\ \quad {P_n}^X(f) = (1-\tilde{D}^2){P}_n(f),
\end{align}
where $\tilde{D} = e^{-i2\pi  f L}$.
The resulting sensitivity under such simplification reads
\begin{align}
{h_c}^X(f) = \sqrt{f\frac{{P_n}^X(f)}{\mathcal{R}^X(f)} }= \sqrt{f\frac{{P_n}(f)}{\mathcal{R}(f)}} = h_c(f).
\end{align}
This confirms that the sensitivity in Eq.~\eqref{eq:18} based on the simplified Michelson-type observable Eq. (\ref{ep:7}) can serve as a good assessment of the sensitivity considering TDI data combinations. Detailed analysis incorporating more realistic orbital dynamics and unequal-arm constellations is left for future studies.

Another important aspect to consider is the tidal disturbances from the gravitational environment in space, including disturbances from Earth's geopotentials, third body perturbations from the Moon, the Sun, etc., which cannot be ignored below 0.1 mHz. Such disturbances are intrinsically deterministic and could be precisely modeled, fitted, and subtracted based on the existing Earth's and Lunar global gravity field models (e.g., from GRACE \citep{tapley2002grace}, GRACE-FO \citep{kornfeld2019grace}, and GRAIL \citep{zuber2013gravity} missions) and planetary ephemeris. Such data processing technique was already routinely and successfully applied in SLR and planetary geodesy \citep{combrinck2010satellite} missions. The development of such a detailed noise subtraction pipeline, while essential for future mission concept studies, involves extensive simulation-based data analysis research that falls outside the scope of this paper. Therefore, we proceed under the assumption that these deterministic signals can be effectively modeled and removed, ignoring their residual effects in the following sensitivity analysis.

Thus, for our idealized and simplified model, four parameters have significant impacts on the sensitivity: $P_{\text{OMS}}$, $P_{\text{ACC}}$, the arm length of the antenna $L$, and the orbital frequency $f_0$. Table~\ref{tab:1} presents the optimal design parameters of various geocentric GW antennas. Based on these parameters, we now analyze the sensitivity curves that demonstrate the impact of orbital motion.

This analysis accounts for the coupling between the detector’s periodic orbital motion and the incoming GW signal. When the GW frequency ($f_{\mathrm{GW}}$) matches specific harmonics of the detector's orbital frequency ($f_{\mathrm{GW}} = n f_0$, for $n=1,2,3$), an orbital resonance occurs, amplifying the detector's response. To quantify the fundamental impact of this effect on detection capability, we plot the characteristic strain sensitivity curves separately for the inspiral and merger phases.

\begin{figure}[ht]
    \centering
    \includegraphics[width=0.50\textwidth]{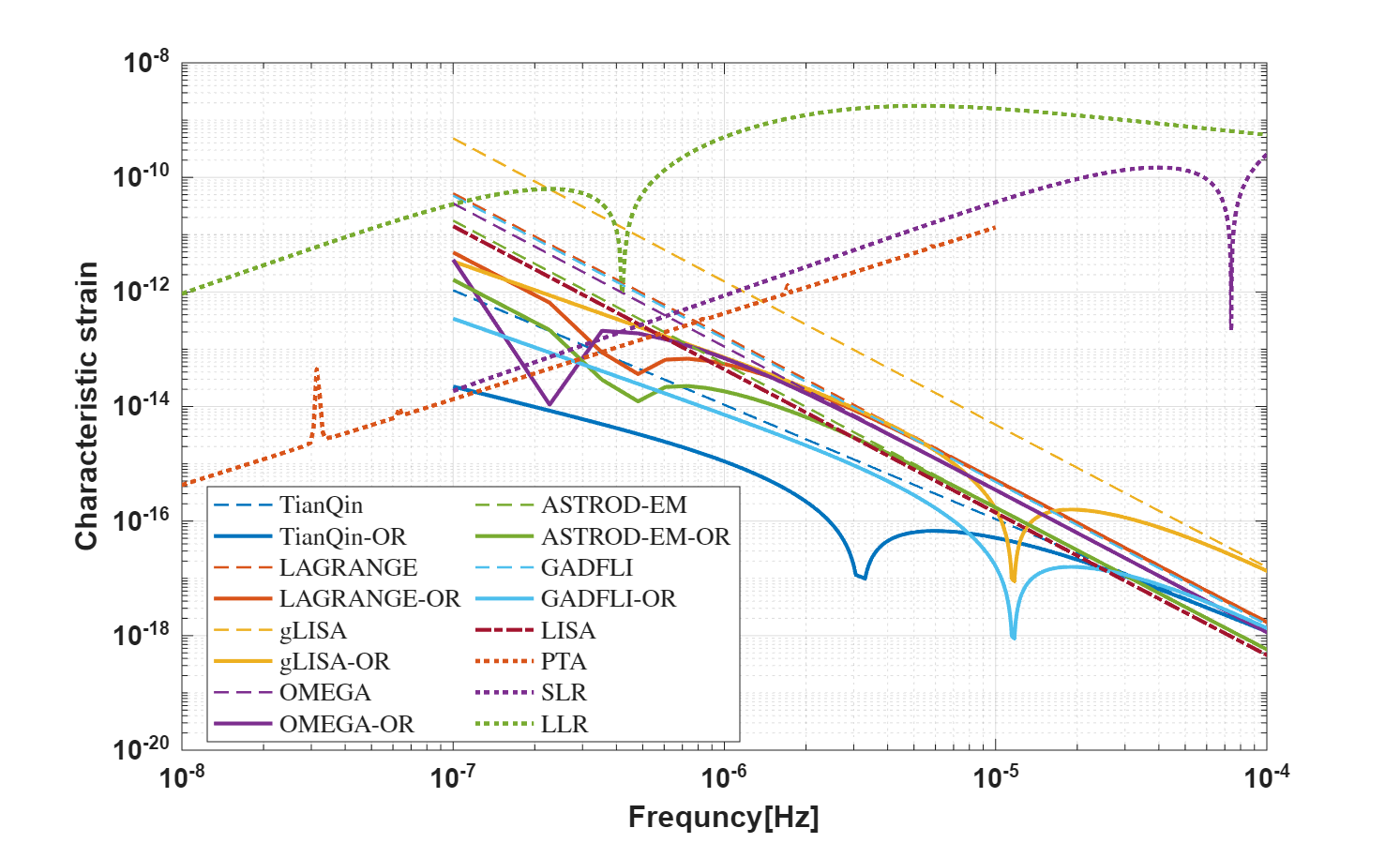}
    \caption{Characteristic strain sensitivity of geocentric GW antennas for the \textbf{inspiral phase}. Curves labeled ``-OR'' (e.g., ``TianQin-OR'') include the full orbital motion and resonant response, while unlabeled curves (e.g., ``TianQin'') represent the non-resonant baseline. The comparison directly quantifies the enhancement from orbital dynamics for inspiral signals.}
    \label{fig:3}
\end{figure}
\begin{figure}[ht]
    \centering
    \includegraphics[width=0.50\textwidth]{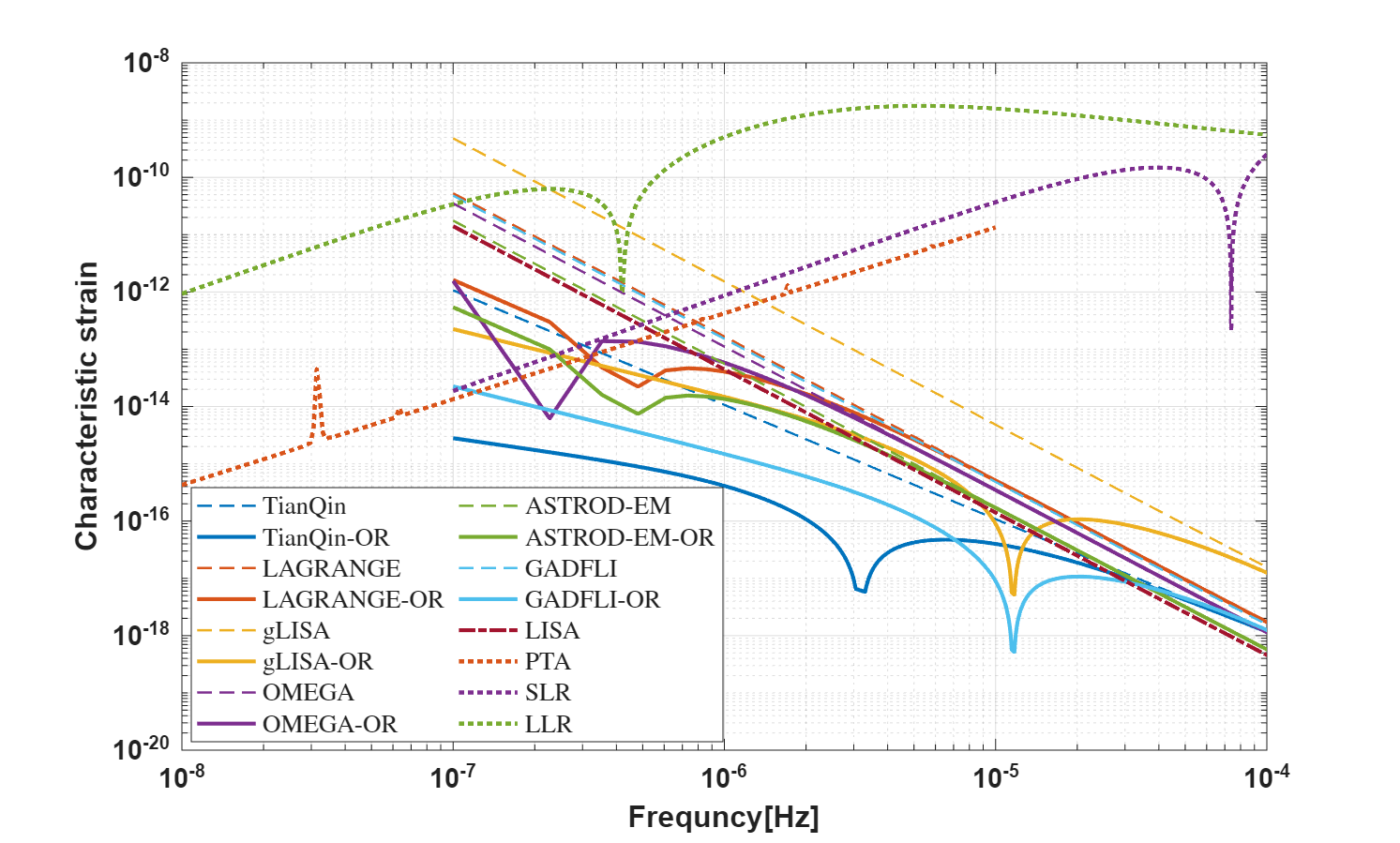}
    \caption{Characteristic strain sensitivity of the same detectors for the \textbf{merger phase}. The curve conventions and the origin of the resonant peak are identical to those in Fig.~\ref{fig:3}. The key difference lies in the signal amplitude: the merger signal is intrinsically stronger, leading to generally better sensitivity (lower curves) across the frequency band compared to the inspiral phase.}
    \label{fig:4}
\end{figure}

Figure~\ref{fig:3} shows the sensitivity for the inspiral phase based on the parameters in Table~\ref{tab:1}. For all detectors, the curves including orbital resonance (``-OR'') exhibit a remarkable sensitivity improvement (lower characteristic strain) near the specific resonant frequencies, with gains reaching one to two orders of magnitude. This result unambiguously confirms that fully accounting for orbital dynamics is crucial and cannot be neglected when evaluating the performance of such detectors, especially in the low-frequency ($\mu$Hz) band. The traditional simplified treatment (non-resonant baseline curves) would severely underestimate the detector's true potential. A notable feature is that the resonant response for each detector manifests as only one dominant peak in the frequency domain. This occurs because, within the observation band, the responses driven by the $f_{\mathrm{GW}} = f_0$ and $f_{\mathrm{GW}} = 3f_0$ resonance conditions overlap at the same frequency point in the Fourier domain. The resonance at $f_{\mathrm{GW}} = 2f_0$ corresponds to a much lower frequency that falls within a noise-dominated regime, rendering its contribution invisible on the sensitivity curve. It should be noted that in the time domain, the responses driven by the $f_0$ and $3f_0$ harmonics occur at different epochs as the GW frequency evolves through the corresponding resonant points, making them in principle distinguishable, as illustrated in Figure~\ref{fig：1}.

Figure~\ref{fig:4} presents the sensitivity for the merger phase. The physical origins of the resonance peak are identical to those in Fig.~\ref{fig:3}. A key observation is that the detector exhibits higher sensitivity during the merger stage than during the inspiral stage. However, this higher sensitivity is applicable only to a narrow frequency band of the full GW signal. In contrast, the inspiral-based sensitivity in Fig.~\ref{fig:3} covers a much broader frequency range, reflecting the signal's long-term evolution. Taken together, Figs.~\ref{fig:3} and~\ref{fig:4} demonstrate that the orbital resonance consistently enhances the detector response across different signal regimes. The final detection prospects are thus jointly determined by these complementary contributions: the sustained frequency evolution of the inspiral and the transient but stronger response of the merger.

\begin{figure}[ht]
    \centering
    \includegraphics[width=0.50\textwidth]{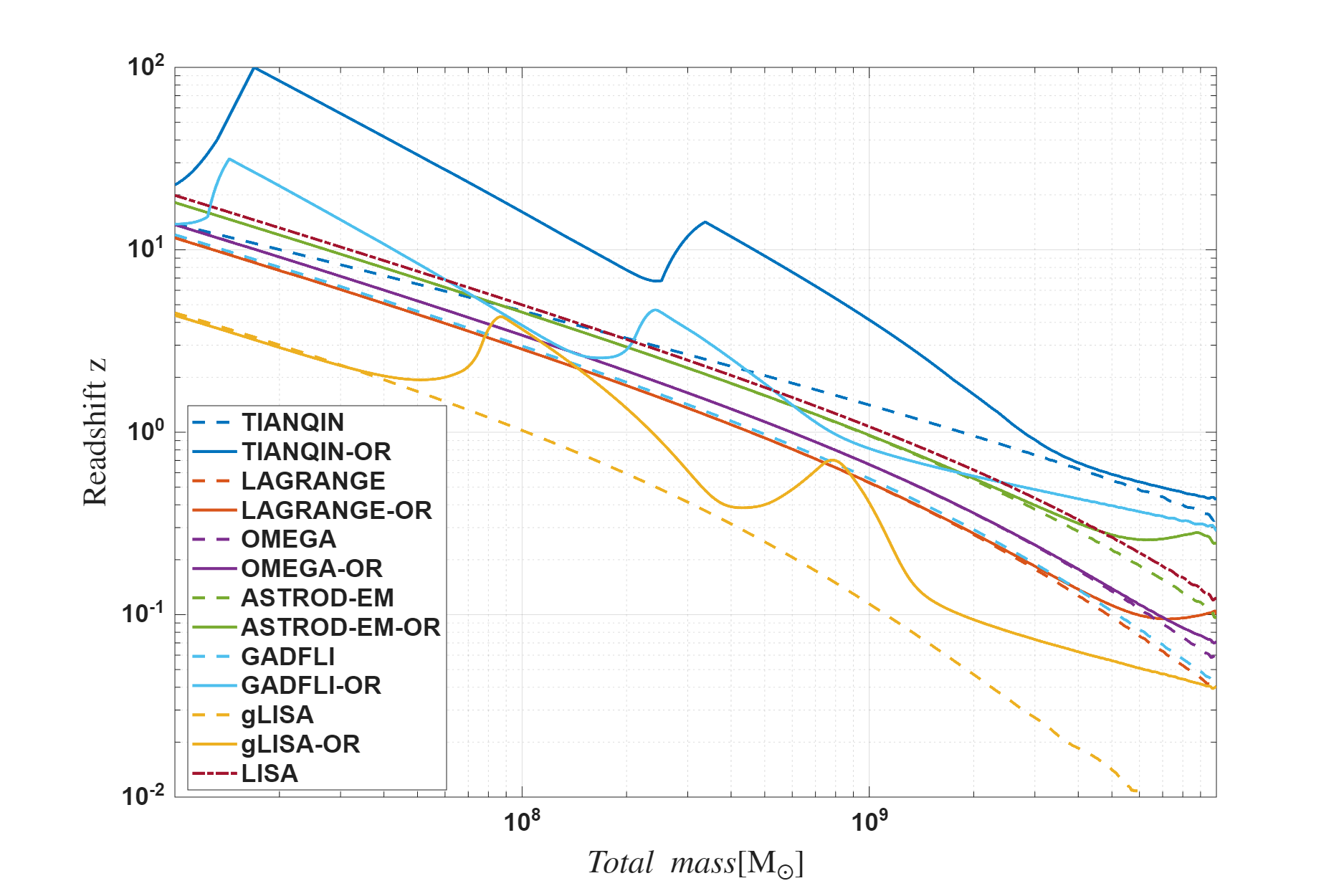}
    \caption{Detectable region in the SMBHB total mass--redshift parameter space. Solid curves: Detection threshold (SNR=10) calculated using the full IMR waveform and including orbital resonance effects. Dashed curves: Threshold without resonance effects. A three-month observation period is assumed, matching typical mission science windows.}
    \label{fig:5}
\end{figure}

While the frequency-domain sensitivity curves in Figs.~\ref{fig:3} and~\ref{fig:4} quantify the instantaneous response, assessing the actual detectability of specific sources requires integrating this response over the observed signal evolution, which depends on source parameters (like mass and redshift) and observation time. To translate the resonant enhancement into a concrete scientific forecast, we evaluate its impact on the detectable population of SMBHBs. This leads to the mass-redshift parameter space plot shown in Fig.~\ref{fig:5}, which integrates the complete IMR waveform, a realistic three-month observation window, and a signal-to-noise ratio threshold of 10 to map out the regions where each detector can successfully identify sources.

The results in Fig.~\ref{fig:5} reveal how the resonance enhancement depends critically on the detector's orbital frequency. For detectors with higher orbital frequencies ($f_0 \sim 10^{-5}$ Hz), such as TianQin, gLISA, and GADFLI, the resonance effect substantially extends the detectable region, particularly for SMBHBs in the mass range of $10^7$--$10^9 M_{\odot}$. A clear double-peak structure emerges in the solid curves (with resonance) when compared to the dashed baselines. This structure arises because a fixed three-month observation samples a limited segment of the signal's frequency evolution. As the total mass increases, this sampled segment shifts to lower frequencies. The two peaks correspond precisely to this segment successively crossing the two strongest orbital resonance bands at $3f_0$ and $f_0$, creating two local maxima in detectability. Conversely, for missions with very low orbital frequencies ($f_0 \sim 10^{-7}$ Hz), such as OMEGA, LAGRANGE, and ASTROD-EM, the corresponding resonance bands lie in an ultra-low-frequency regime where the detector noise is prohibitively high. Consequently, while a minor improvement might be visible for the most massive sources ($M_{\text{Total}} \gtrsim 10^{10} M_{\odot}$), the resonance effect provides negligible improvement to the detectable parameter space for these detectors across most of the considered mass range.

\begin{table*}
\centering
\caption{Parameters of geocentric GW antennas}
\begin{tabular}{lcccc}
\hline
\hline
Missions & $\sqrt{P_{\mathrm{ACC}}}$ & $\sqrt{P_{\mathrm{OMS}}}$ & $L$ & $f_0$ \\
 & ($\mathrm{fm/s^2}/\mathrm{Hz}^{1/2}$) & ($\mathrm{pm}/\mathrm{Hz}^{1/2}$) & (Gm) & ($\mu\mathrm{Hz}$) \\
\hline
ASTROD-EM \citep{ni2016gravitational} & $1\cdot\sqrt{1 + \bigl(0.4\,\mathrm{mHz}/f\bigr)^2}$ & 1 & 0.66 & 0.42227 \\
GANDFLI \citep{mcwilliams2011geostationary} & $0.3\cdot\sqrt{1 + \bigl(0.4\,\mathrm{mHz}/f\bigr)^2}$ & 1 & 0.073 & 11.606 \\
gLISA/GEOGRAWI \citep{tinto2011geostationary} & $3\cdot\sqrt{1 + \bigl(0.4\,\mathrm{mHz}/f\bigr)^2}$ & 0.3 & 0.073 & 11.606 \\
LAGRANGE \citep{conklin2011lagrange} & $3\cdot\sqrt{1 + \bigl(0.4\,\mathrm{mHz}/f\bigr)^2}$ & 5 & 0.66 & 0.42227 \\
OMEGA \citep{hellings2011low} & $3\cdot\sqrt{1 + \bigl(0.4\,\mathrm{mHz}/f\bigr)^2}$ & 5 & 1 & 0.22905 \\
TianQin \citep{huang2020science} & $1\cdot\sqrt{1 + 0.1\,\mathrm{mHz}/f}$ & 1 & 0.11 & 3.1775 \\
\hline
\label{tab:1}
\end{tabular}
\end{table*}

\section{Conclusions} 
Previous studies have demonstrated the importance of orbital resonance effects for GW detection in geocentric laser ranging systems, including SLR and LLR missions. However, orbital modulations and resonance phenomena in the $\mu$Hz band remain unexplored for geocentric LISA-like GW antennas. Our analysis reveals that these long-overlooked effects will strikingly enhance detector sensitivities by 1-2 orders of magnitude in the $\mu$Hz band compared to those obtained from traditional response models (Figure \ref{fig:3} and Figure \ref{fig:4}). Crucially, this enhancement is most pronounced for detectors with orbital frequencies around $10^{-5}$ Hz, such as TianQin, gLISA, and GADFLI, whose sensitivity curves exhibit a well-defined resonant peak. As shown in Figure \ref{fig:5}, this sensitivity gain would expand the detectable parameter space for coalescing SMBHBs and significantly extend the observational horizon of these geocentric antennas. In particular, for the aforementioned detectors, the inclusion of orbital resonance leads to a distinctive double-peak structure in the detectable region within the mass range of $10^7$–$10^9 M_{\odot}$, corresponding to the sequential crossing of the $3f_0$ and $f_0$ resonance bands during a finite observation window. 
Crucially, such $\mu$Hz improvement could provide continuous coverage of late-stage SMBHB dynamics, synergizing with PTA observations of the same systems decades earlier in their nanohertz inspiral phase and enabling unprecedented joint tests.
Importantly, as feasible near-term precursors for $\mu$Hz astronomy avoiding the technical and budgetary constraints, this resonance-driven approach transforms geocentric LISA-like antennas (e.g. $\mu$Ares, LISAmax, ASTROD-GW, Super-ASTROD) into indispensable platforms for bridging the GW spectrum, providing an indispensable piece in the overall picture of black-hole growth and galaxy evolution. Our results indicate that for detectors with much lower orbital frequencies (e.g., OMEGA, LAGRANGE, ASTROD-EM), the resonance enhancement is largely overshadowed by the elevated noise floor in the corresponding ultra-low-frequency band, underscoring that the benefits of orbital resonance are highly dependent on the specific detector design.

The present study employs a simplified model assuming circular orbits and equilateral triangular constellations to establish a unified framework for comparing the resonance-enhanced sensitivities of various geocentric GW missions, such as TianQin, gLISA, and GADFLI that will operate in different orbital environments. This idealized configuration enables us to isolate and quantify the intrinsic sensitivity improvement due to orbital resonance, disentangling it from mission-specific complexities. 
While the model does not account for arm-length variations induced by injection errors or gravitational perturbations, such as those from Earth’s multipole moments or third-body interactions, these simplifications are both necessary and conventional in analytical studies aiming to extract physical insight and enable consistent cross-mission comparisons.
Future work will extend this analytical study through high-fidelity, mission-specific numerical simulations incorporating realistic orbital dynamics and environmental perturbations. A key focus will be the development of advanced data processing techniques, including precision gravity-field modeling and data-driven disturbance subtraction methods—building on well-established approaches used in satellite laser ranging and geodesy—to mitigate deterministic noises. Furthermore, we aim to incorporate full TDI into the resonance-enhanced responses, perform systematic surveys of the parameter space for coalescing SMBHBs, and assess the detectability of stochastic GW backgrounds for resonance–enhanced geocentric antennas.

\begin{acknowledgments}
This work is supported by the National Key Research and Development Program of China No. SQ2024YFC220046 and the International Partnership Program of the Chinese Academy of Sciences, No. 025GJHZ2023106GC.
\end{acknowledgments}

\appendix
\section{Orbital expansion for small eccentricity}\label{AppendixA}

When the orbital eccentricity is small, one can expand the orbit equations in the eccentricity as a small parameter. 
Using the cylindrical coordinate system introduced in Sec. \ref{scheme}, the Keplerian equations of motion for each spacecraft read
\begin{align}
    \ddot{R}_n - R_n\dot{\Phi}_n^2 + \frac{\mu}{(R_n^2+Z_n^2)^{3/2}}\,R_n &= f_{R_n}, \label{A1}\\
    2\dot{R}_n\dot{\Phi}_n + R_n\ddot{\Phi}_n &= f_{\Phi_n}. \label{A2}
\end{align}
As in Sec. \ref{scheme}, the motion in the $z$ direction has a negligible effect on the ranging measurements and is therefore neglected. 
Substituting Eqs.~\eqref{eq:3}--\eqref{eq:4} into Eqs.~\eqref{A1}--\eqref{A2} yields, to leading order in the small parameter $\epsilon$,
\begin{align}
    \ddot{F}_n + 2\frac{\dot{r}_n}{r_n}\dot{F}_n - 2\dot{\phi}_n\dot{G}_n - 3\frac{\mu}{r_n^3}F_n &= \frac{f_{R_n}}{\epsilon\,r_n} + O(\epsilon^2), \label{A3}\\
    \ddot{G}_n + 2\frac{\dot{r}_n}{r_n}\dot{G}_n + 2\dot{\phi}_n\dot{F}_n &= \frac{f_{\Phi_n}}{\epsilon\,r_n} + O(\epsilon^2).\label{A4}
\end{align}
Since the eccentricity $e$ is a small parameter, the coupled equations~\eqref{A3}--\eqref{A4} can be expanded as a power series in $e$:
\begin{align}
    F_n &= F_n^{(0)} + F_n^{(1)} e + \cdots, \\
    G_n &= G_n^{(0)} + G_n^{(1)} e + \cdots, \\
    F_n^{(k)} &= \frac{1}{k!} \left. \frac{\partial^k F_n}{\partial e^k} \right|_{e=0}, \\
    G_n^{(k)} &= \frac{1}{k!} \left. \frac{\partial^k G_n}{\partial e^k} \right|_{e=0}.
\end{align}
At each order in $e$, one obtains a corresponding set of differential equations. For the zeroth and first orders, they take the following forms:
\begin{align}
    \ddot{F}_n^{(0)} - 2\omega_0 \dot{G}_n^{(0)} - 3\omega_0^2 F_n^{(0)} &= 
    \left. \frac{f_{R_n}}{\epsilon r_n} \right|_{e=0}, \label{A5}\\
    \ddot{G}_n^{(0)} + 2\omega_0 \dot{F}_n^{(0)} &= 
    \left. \frac{f_{\Phi_n}}{\epsilon r_n} \right|_{e=0}, \label{A6}\\
    \ddot{F}_n^{(1)} - 2\omega_0 \dot{G}_n^{(1)} - 3\omega_0^2 F_n^{(1)} 
    + 2\omega_0 \sin(\omega_0 t)\, \dot{F}_n^{(1)} &=
    \left. \frac{\partial}{\partial e} \left( \frac{f_{R_n}}{\epsilon r_n} \right) \right|_{e=0}
    + 4\omega_0 \cos(\omega_0 t)\, \dot{G}_n^{(0)}
    + 9\omega_0^2 \cos(\omega_0 t)\, F_n^{(0)}, \label{A7}\\
    \ddot{G}_n^{(1)} + 2\omega_0 \dot{F}_n^{(1)} 
    + 2\omega_0 \sin(\omega_0 t)\, \dot{G}_n^{(1)} &=
    \left. \frac{\partial}{\partial e} \left( \frac{f_{\Phi_n}}{\epsilon r_n} \right) \right|_{e=0}
    - 4\omega_0 \cos(\omega_0 t)\, \dot{F}_n^{(0)},\\
    &\cdots
\end{align}

Clearly, Eqs.~\eqref{A5}--\eqref{A6} correspond to the circular-orbit solution. 
Reference~\cite{mashhoon1978tidal} analyzed the time-domain response of a binary system to a monochromatic GW, showing that the amplitudes of $F_n^{(k)}$ and $G_n^{(k)}$ remain of comparable magnitude across different orders $k$.
Therefore, when $e$ is small, the circular-orbit component $F_n^{(0)}$ and $G_n^{(0)}$ dominates, while the contribution from the $k$th-order correction $F_n^{(k)}$ and $G_n^{(k)}$ is smaller by a factor of $O(e^k)$.

\bibliography{GW_orbital_resonance}{}
\bibliographystyle{aasjournalv7}



\end{document}